\documentclass[11pt,aps,prc,superscriptaddress,showpacs]{revtex4}

\usepackage[dvips]{graphicx}
\usepackage{amsmath,amssymb}

\newcommand{\vj}{\mbox{\boldmath $j$}}
\newcommand{\vD}{\mbox{\boldmath $D$}}
\newcommand{\vp}{\mbox{\boldmath $p$}}
\newcommand{\va}{\mbox{\boldmath $a$}}
\newcommand{\vk}{\mbox{\boldmath $k$}}

\newcommand{\psla}{\ooalign{\hfil/\hfil\crcr{p}}}
\newcommand{\ksla}{\ooalign{\hfil/\hfil\crcr{k}}}
\newcommand{\qsla}{\ooalign{\hfil/\hfil\crcr{q}}}
\newcommand{\delsla}{\mbox{\ooalign{\hfil/\hfil\crcr{$\partial$}}}}

\newcommand{\fpisq}{\mbox{$f_{\pi}^2$}}
\newcommand{\mpi}{\mbox{$m_{\pi}$}}
\newcommand{\mpisq}{\mbox{$m_{\pi}^2$}}

\newcommand{\eq}{\label}

\newcommand{\psibar}{\mbox{$\overline{\psi}$}}
\newcommand{\Pitild}{\mbox{${\tilde{\Pi}}$}}
\newcommand{\vPitild}{\mbox{${\tilde{\boldmath \Pi}}$}}

\newcommand{\epsi}{\mbox{$\varepsilon$}}

\def\sigm{\mbox{$\langle \sigma \rangle$}}
\def\gs{\mbox{$g_\sigma$}}
\def\gv{\mbox{$g_\omega$}}
\def\ms{\mbox{$m_\sigma~$}}
\def\mv{\mbox{$m_\omega~$}}

\begin{document}

\title{
Nuclear Electromagnetic Current 
in the Relativistic Approach \\
with the Momentum-Dependent Selfenergies}

\author{Tomoyuki~Maruyama}
\affiliation{College of Bioresource Sciences,
Nihon University, Fujisawa 252-8510, Japan}
\affiliation{Advanced Science Research Center, \\
Japan Atomic Energy Research Institute,Tokai 319-1195, Japan}

\author{Satoshi Chiba}
\affiliation{Advanced Science Research Center, \\
Japan Atomic Energy Research Institute,Tokai 319-1195, Japan}

\begin{abstract}
We define the new description of the electromagnetic current to hold 
the current conservation in the momentum-dependent Dirac fields from the
Ward Takahashi identity.
To describe the momentum-dependence we solve the relativistic
Hartree-Fock approximation by using the one-pion exchange.
In addition we discuss on contribution from the one-pion exchange current and
the core polarization.
It is shown that this current can reduce 
the convection current in the isovector case, 
whose value has been too big due to the small effective mass 
in the usual relativistic Hartree approximation. 
\end{abstract}


\maketitle

%\vfil
%\eject
%\newpage

\section{Introduction}

The past decades have seen many successes in the relativistic 
treatment of the nuclear many-body problem.
The relativistic framework has big advantages in several aspects \cite{Serot}:
a useful Dirac phenomenology for the description of nucleon-nucleus
scattering \cite{Hama,Tjon}, the natural incorporation of the spin-orbit force 
\cite{Serot} and the saturation properties of nuclear matter 
in the microscopic treatment
with the Dirac Brueckner Hartree-Fock (DBHF) approach \cite{DBHF}. 

These results conclude that there are large attractive scalar and 
repulsive vector-fields, and that the nucleon effective mass
is very small in the medium.
However this small effective mass leads to small Fermi velocity, 
which makes some troubles in the nuclear
properties: too big magnetic-moment \cite{magmom} and 
too big excitation energy of
the isoscalar giant quadrupole resonance (ISGQR) state \cite{Nishi}. 
As for the isoscalar magnetic-moment, this enhancement is cancelled 
by the Ring-Diagram contribution \cite{Ring};
this relation is completely realized by the gauge invariance \cite{Bentz}. 
As for the isovector one, however, this contribution does not plays 
a significant role because 
the symmetry force is not efficiently large.

In this subject most of people believed that 
the momentum dependence of the Dirac fields is negligible 
in the low energy region, particularly below the Fermi level.
A momentum dependence of the Schr\"odinger equivalent potential
automatically emerges as a consequence of the Lorentz transformation properties
of the vector-fields without any explicit momentum dependence of the scalar and
vector fields.
In fact, only very small momentum dependence has appeared in 
the relativistic Hartree-Fock (RHF) calculation \cite{RHF,soutome}.

In the high energy region, however, the vector-fields must become very small 
to explain the optical potential of the proton-nucleus elastic scattering 
\cite{Hama,KLW1}, and the transverse flow in the heavy-ion collisions 
\cite{TOMO1}.
The momentum-dependent part is not actually small
though it has not been clearly seen in the low energy phenomena.
Furthermore S.~Typel \cite{Typel} introduces the non-local parts
and succeeded to improved nuclear properties.

In the previous paper \cite{tomoGRF} 
we showed that the momentum dependence of the Dirac-fields
is very sensitive to the Fermi velocity though it 
affects little the nuclear equation of state.
In that work
we introduced the one-pion exchange force, which 
produce the dominant contribution of 
the momentum dependence and suppresses the Fermi velocity,
and explain the ISGQR energy.

We can easily imagine that the one-pion exchange force 
largely produce the momentum dependence 
because the interaction range is largest.
Since the momentum-dependent fields break the current conservation, 
we have to define the new current caused by the vertex correction.

In this paper, thus, we investigate the nuclear current using 
the momentum-dependent Dirac fields.
For this purpose we define a new current to hold the 
current conservation in the momentum-dependent Dirac fields, 
and discuss its effect on the the nuclear static current.
In this work we focus only on the convection current which is
sensitive to the Fermi velocity, and omit the spin current.  

In the next section we explain our formalism to make a conserved current
under the momentum-dependent selfenergies.
In Sec. 3 we show our numerical results for the static current in our
formulation.
Then we summarize our work in Sec. 4. 

\newpage

\section{Formalism}

\subsection{Nucleon Propagator}

Now we describe the propagator of nucleon with momentum $p$ 
in the isospin space as follows.
\begin{equation}
S(p) = \left(
 \begin{array}{cc}
 S_p (p) & 0 \\
    0    & S_n (p) 
\end{array}
\right)
\end{equation}
Here we assume the spin isospin-saturated nuclear matter, and define
the proton and neutron propagator as
\begin{equation}
S_N (p) = S_p (p) = S_n (p) .
\end{equation}

The nucleon propagator in the selfenergy $\Sigma$ is 
given by 
\begin{equation} 
S^{-1}_N (p) = \psla - M - \Sigma(p),
\label{prop}
\end{equation} 
where $\Sigma(p)$ has a Lorentz scalar part $U_s$ and a Lorentz vector 
part $U_{\mu}(p)$ as
\begin{equation}
\Sigma(p) = - U_{s}(p) + \gamma^{\mu} U_{\mu}(p).
\end{equation} 
For the future convenience we define the effective mass and the kinetic 
momentum as
\begin{eqnarray}
M^{*}(p) & = & M - U_{s}(p) ,
\nonumber \\
\Pi_{\mu}(p) & = & p_{\mu} - U_{\mu}(p) .
\end{eqnarray}
The single particle energy with momentum {\vp} is obtained as
\begin{eqnarray}
\varepsilon({\vp}) & = & p_0 |_{on-mass-shell}
\nonumber \\
& = & \sqrt{ \Pi^2({\vp}) + M^{*2} } + U_{0}(\vp) . 
\end{eqnarray}
Then the detailed form of the nucleon propagator eq. (\ref{prop}) is
represented by
\begin{equation}
S_N (p) = S_F (p) + S_D (p) 
\end{equation}
with
\begin{eqnarray}
S_F (p) = \{ \not{\Pi}(p) + M^{*}(p) \} 
 \frac{1}{\Pi^2 - M^{*2} + i\delta } 
\\  
S_D (p) = 2 i \pi \{ \not{\Pi}(p) + M^{*}(p) \} 
n(\vp) \theta (p_0)  \delta [ V(p) ] ,
\label{propd}
\end{eqnarray}
where $n(\vp)$ is the momentum distribution, 
and
\begin{equation}
V(p) \equiv \frac{1}{2} \{ {\Pi}^2(p) - M^{*2}(p) \} .
\end{equation}

% \equiv \Theta (p_F -|\vp|)$, where $\Theta$ is a step function.

\subsection{Momentum-Dependent Selfenergies}

We can easily suppose that it is the one-pion exchange force 
which produces the major momentum dependence because the 
interaction range is largest.
In this work, thus, we introduce the momentum dependence
to the Dirac fields due to the one-pion exchange, and 
discuss how the Fock parts affects the nuclear current.  

Along this line
we define a Lagrangian density in the system as
\begin{eqnarray}
{\cal L} & = & {\psibar} ( i \delsla - M ) \psi
+ \frac{1}{2}\partial_{\mu}{\phi_a} \partial^{\mu}{\phi_a}
- \frac{1}{2} m_{\pi}^2 {\phi_a}{\phi_a}
- {\widetilde U} [\sigma] 
+ \frac{1}{2} {m_{\omega}^2} 
\omega_\mu \omega^\mu 
\nonumber \\ 
& & 
\rho_{\mu a} \rho^\mu_a 
+ i \frac{f_{\pi}}{m_{\pi}} 
{\psibar} \gamma_{5} \gamma^{\mu} {\tau}_{a} \psi \partial_{\mu} \phi_{a}
+ \gs \bar{\psi} \psi \sigma
- \gv \bar{\psi} \gamma_\mu \psi \omega^\mu
- \frac{C_v^{IV}}{2M^2} \{ \bar{\psi} \gamma_\mu \tau \psi \}^2
\label{Lag}
\end{eqnarray}
where $\psi$, $\phi$, $\sigma$ and $\omega$ are 
the nucleon, pion, sigma-meson and omega-meson  fields, 
respectively, and the suffix $a$ indicates the isospin component.
In the above expression 
we use the pseudo-vector coupling
form as an interaction between nucleon and pion.
The selfenergy potential of the $\sigma$-field
${\widetilde U} [\sigma]$ is given as
Ref. \cite{TOMO1,K-con}.
\begin{equation}
\widetilde{U} [\sigma] 
= \frac{ \frac{1}{2} {m_{\sigma}^2} \sigma^2 
+ \frac{1}{3} B_\sigma \sigma^3
+ \frac{1}{4} C_\sigma \sigma^4 } 
{ 1 + \frac{1}{2} A_\sigma \sigma^2 } \  .
\eq{sigself}
\end{equation}
The symbols
$m_\pi$, \ms and \mv  are the masses of {$\pi$}-, {$\sigma$}- and
{$\omega$}-mesons, respectively.
In addition, we also introduce the isovector nucleon-nucleon interaction
 into the Lagrangian (\ref{Lag}).  
so as to discuss on the isovector current later.

Next we calculate the nucleon selfenergies.
The nucleon selfenergies are separated into the local part
and the momentum-dependent part as
%\begin{equation}
$U_{\alpha} (p) = U_{\alpha}^{L} + U_{\alpha}^{F} (p),$
where $\alpha = s, \mu$.
%\end{equation}
The {$\sigma$}- and {$\omega$}-meson exchange parts
produce
only very small 
momentum dependence of nucleon selfenergies \cite{RHF,soutome}
as their masses are large.
In fact the RH and RHF approximations
do not give any different results in nuclear matter properties 
after fitting parameters 
of {$\sigma$}- and {$\omega$}-exchanges \cite{RHF}.
%it is very easily confirmed in the limit of 
%$m_{\alpha} \rightarrow \infty$ at the fixed 
%${g_\alpha}/m_{\alpha}$.
On the other hand the one-pion exchange force is a long range 
one and makes for a large momentum dependence while it does not
contribute to the local part in the spin-saturated system.
Subsequently we make the local part by RH
of the {$\sigma$}- and {$\omega$}-meson exchanges, and
the momentum-dependent part by RHF of the pion exchange, 
and thus we omit the kinetic energy part mesons except pion in eq.(\ref{Lag}).
This method is shown in Ref.\cite{KLW1} to
keep the self-consistency within the RHF framework.

In this model the local part of the selfenergies are given as
\begin{eqnarray}
U_{s}^{L} & = & \gs \sigm
\\
U_{\mu}^{L} & = & \delta_{0 \mu} \frac{\gv^2}{m_{\omega}^2} 
\rho_H
\end{eqnarray}
where $\sigm$ is the scalar mean-field obtained as
\begin{equation}
\frac{\partial}{\partial \sigm} {\tilde U} [\sigm]
= \gs \rho_s  
\end{equation}
In the above equations the scalar density $\rho_s$
and the vector Hartree density $\rho_H$ are given by
\begin{eqnarray}
\rho_s & = & 4	\int \frac{{\rm d}^3 {\vp}}{(2 \pi)^3}  
n({\vp}) 
\frac{M_{\alpha}^* (p)}{ \Pitild_0 (p) }   ,
\\
\rho_H & = & 4	\int \frac{{\rm d}^3 {\vp}}{(2 \pi)^3}  
n({\vp}) 
\frac{\Pi_0 (p)}{ \Pitild_0 (p)}   ,
\label{rhos}
\end{eqnarray}
where $n({\vp})$ is the momentum-distribution,
and  $\Pitild_{\mu} (p)$ is defined by
\begin{equation} 
\Pitild_{\mu} (p) = \frac{1}{2} \frac{\partial}{\partial p^{\mu}}
[ \Pi^2 (p) - {M^*}^2 (p) ]
\end{equation}
%%deleted on Oct 4, 1999
%Please note that the Hartree density $\rho_H$ is not equivalent
%to the baryon density $\rho_B$ when the selfenergies depends
%on the four momentum.

As a next step we define
the momentum-dependent parts of the selfenergies as
the Fock parts with the one-pion exchange. 
When using the pseudo vector (PV) coupling 
the Fock parts do not become zero
at the infinite limit of the momentum $|\vp|$.
One usually erases these contributions by introducing 
the cut-off parameter.
In this work, instead of that, 
we subtract these contributions from the momentum-dependent
parts (these contributions 
can be renormalized into the Hartree parts):
$U_\alpha \rightarrow U_\alpha - U_\alpha (p \rightarrow \infty)$.
Thus we obtain the momentum-dependent parts of the selfenergies
as
\begin{eqnarray}
U_{s}^{F} (p) & = &
\frac{3 \fpisq}{2}
\int \frac{d^3 \vk}{(2 \pi)^3} 
n({\vk}) 
\frac{M^* (k)}{ \Pitild_0 (k) } \Delta_{\pi} (p-k)   ,
\label{USPV}
\\
U_{\mu}^{F} (p) & = & 
- \frac{3 \fpisq}{2}
\int \frac{d^3 \vk}{(2 \pi)^3} 
n({\vk}) 
\frac{\Pi_{\mu} (k)}{ \Pitild_0 (k) } \Delta_{\pi} (p-k)  ,
\label{UVPV}
\end{eqnarray}
where the ${\Delta_\pi}(q)$ is the pion propagator defined as
\begin{equation}
{\Delta_\pi}(q) = \frac{1}{q^2 - \mpisq} .
\end{equation}

In the above vector selfenergies
we omit the tensor-coupling part  
involving $[ \Pi(k) \cdot  (p-k)] (p-k)_{\mu}$. 
This term is very small if the selfenergy
is independent of momentum \cite{RHF}, and
their momentum dependence is actually very small 
as shown later.

\bigskip

\subsection{One-Body Current Operator}

If the selfenergy has a momentum dependence, the current operator
must be also changed to satisfy the current conservation.
We then define the current vertex $\Gamma^{\mu}(p+q,p)$ as  
\begin{equation}
\Gamma^{\mu}(p+q,p) = \gamma^\mu + \Lambda^\mu(p+q,p).
\end{equation}
The Ward-Takahashi (WT) identity  gives the following relation 
about the current vertex 
\begin{equation}
S_N (p+q) q^{\mu} \Gamma_{\mu}  S_N (p) =  - S_N (p+q) + S_N (p) .
\label{WTI}
\end{equation}
This expression is rewritten as
\begin{equation}
q^{\mu} \Gamma_{\mu}  =  S^{-1}_N (p+q) - S^{-1}_N (p) .
\label{WTGam}
\end{equation}
Substituting eq.(\ref{prop}) into the above equation (\ref{WTGam}), 
the density-dependent vertex correction $\Lambda^\mu$  is obtained as
\begin{equation}
q_\mu \Lambda^{\mu}(p+q,p) = - \Sigma(p+q) + \Sigma(p).
\label{con-c}
\end{equation}

In this work we restrict our discussion on the convection current,
not on the spin current.
In Appendix A we show an approximate method to derive 
the vertex correction  from the above WT identity
though the anomalous current, which is proportional to 
$\sigma_{\mu \nu} q^{\nu}$, still has an ambiguity
because the WT identity cannot make any condition for it.

When we use the one-pion exchange force,
we have to take into account the meson-exchange current.
In Appendix B we show that
our approximate formulation is also satisfied
in the electro-magnetic current by including the one-pion
exchange current, and the electromagnetic current operator is
proportional to $(1 + \tau_3)/2$, namely only protons contribute to
the convection current.

In this work we focus on the static current in the nuclear matter, then 
we need to get the zero limit of the momentum transfer $q$.
In this limit  the vertex correction becomes
\begin{equation}
\Lambda^{\mu}(p) = \lim_{q \rightarrow 0} \Lambda^{\mu} (p+q,p)
= - \frac{\partial}{\partial p_\mu} \Sigma (p) .
\end{equation}
Using the above vertex correction, the current density of the whole system
is given as
\begin{eqnarray}
\vj_\mu & = & \int \frac{d^4 p}{(2 \pi)^4} {\rm Tr} 
\{ \frac{1+\tau_3}{2} 
(\gamma_\mu - \frac{\partial \Sigma}{\partial p_\mu} ) S (p) \}
\nonumber \\
 & = & \int \frac{d^4 p}{(2 \pi)^3} n^{(p)} ({\vp}) 
\{ \Pi_{\mu}(p) - \Pi^\nu (p) \frac{\partial U_\nu (p)}{\partial p_\mu}
+ M^* (p) \frac{\partial U_s (p)}{\partial p_\mu} \}
\delta[V(p)]
\nonumber \\
 & = & \int \frac{d^3 p}{(2 \pi)^3} n^{(p)} ({\vp}) 
\left . \frac{\Pitild_{\mu}(p)}{\Pitild_{0}(p)} \right|_{p_0 = \epsi(\vp)} ,
\end{eqnarray}
where $n^{(p)}$ is the Fermi distribution function for proton, and
$\Pitild_{\mu}$ is defined by
\begin{eqnarray}
\Pitild_{\mu} (p) & \equiv & \frac{\partial}{\partial p^\mu} V(p)
\nonumber \\
 & = & 
\frac{1}{2} \frac{\partial}{\partial p^\mu} \{ \Pi^{2}(p) - \ M^{*2}(p) \} .
\end{eqnarray}

Let us consider the one-particle state on the Fermi surface.
The space current density contributed from this nucleon can be written as
\begin{equation}
\vj = \frac{{\bf \vPitild}(p)}{\Pitild_{0}(p)} |_{|{\vp}| = p_F}
= \vD_{\vp} \varepsilon ({\vp}) |_{|{\vp}| = p_F},
\end{equation}
where the total derivative ${\vD}_{\bf p}$ is defined on the on-mass-shell
condition: $p_0 = \epsi ({\vp})$.
The above equation completely agrees with that derived by the semi-classical
way \cite{KLW1}.

In the non-relativistic framework the effective mass  is defined by
\begin{equation}
M^{*}_{L} = 
(2 \frac{d}{d \vp^2} \epsi({\vp}) )^{-1}
|_{|{\vp}| = p_{F}},
\end{equation}
which is so called the '{\bf Landau mass}'.
Then the above spatial current density is
\begin{equation}
{\vj} = \frac{{\vp}_F}{M^{*}_{L}}.
\end{equation}
In our case including the momentum-dependent Dirac fields, the value of 
the Landau mass $M^{*}_{L}$ cannot be uniquely determined from 
the relativistic effective mass $M^{*}$ while in the Hartree approximation 
the Landau mass becomes $M^{*}_{L} = \Pi_0({\bf p}_F)$.

\bigskip

\subsection{Core Polarization Current}

As for the actual nuclear current observed by experiments
the core polarization also plays an important role.
Of course this effect cancels contribution of the effective mass
in the isoscalar case \cite{BM}.

Here we should consider a system that 
one valence nucleon populates a state on
the Fermi surface of the saturated nuclear matter.
In this system  the momentum-distribution can be described as
\begin{equation}
n(\vp,\tau) = n_{0}({\vp}) + \frac{1}{4} \Delta n ({\vp},\tau), 
\label{p-var}
\end{equation}
where $n_0 ({\vp}) = \theta (p_F - |\vp|)$ shows 
the usual Fermi distribution
with the Fermi momentum $p_F$, and 
$\Delta n({\vp}) \propto \delta (|\vp| - p_F)$
indicates the valence nucleon part.
The suffix $\tau$ decades the isospin for the valence nucleon,

The valence nucleon varies the selfenergies of nucleons below Fermi surface
from that at the saturated matter as
\begin{equation}
U_{\alpha} (p) \rightarrow U_{\alpha} (p) + \Delta U_{\alpha} (p) .
\label{U-var}
\end{equation}
In addition the function $V(p)$ is also varied as
\begin{equation}
V(p) = V_0 (p) + \Delta V(p)
\label{V-var}
\end{equation}
with
\begin{equation}
\Delta V(p) = - \Pi^{\mu} \Delta U_{\mu}
+ M^{*} \Delta U_{s} .
\label{delV}
\end{equation}

The current density is described with the following expression.
\begin{equation}
j_{tot}^{\mu} =  -\sum_{\tau = \pm 1} \int \frac{d^4 p}{(2 \pi)^3} 
f(p,\tau) \frac{\partial V(p)}{\partial p^\mu } ,
\end{equation}
where $f(p,\tau)$ is 
the four-dimensional momentum-distribution for nucleon with isospin $\tau$,
which is given as
\begin{eqnarray}
f(p,\tau) &=& n(\vp,\tau) \delta ( p_0 - \epsi_{\vp} )
\\
            &=& \frac{1}{\Pitild_0 (p)} n(\vp,\tau) \delta [ V(p) ] \theta (p_0) .
\end{eqnarray} 
The variation along eqs. (\ref{p-var}) $-$ (\ref{delV}) leads to the
above
four dimensional  momentum-distribution $f(p,\tau)$ as
\begin{equation}
f(p,\tau) = f_0 (p,\tau) + \Delta f(p,\tau)
\end{equation}
with
\begin{eqnarray}
f_0 (p,\tau) &=& n_0(\vp) \delta [ V_0(p) ] \theta (p_0) ,
\\
\Delta f(\vp,\tau) &=& \Delta n(\vp,\tau) \delta [ V_0(p) ] \theta (p_0)
+ n_0(\vp) \left \{ \frac{\partial \delta[V(p)]}{\partial V} \right \}_{V = V_0} \Delta V(p,\tau) \theta (p_0) .
\end{eqnarray} 
The first term comes from the valence nucleon, and the second one from
the core polarization.

The total current density is given as
\begin{eqnarray}
j_{tot}^{\mu} &=& \int \frac{d^4 p}{(2 \pi)^3} f (p) \delta [{V(p)}] 
\frac{\partial V(p)}{\partial p_{\mu}} 
 \\
&=&  \delta^{\mu}_0 \rho_B  +j_{val}^{\mu} + j_{cor}^{\mu} . 
\end{eqnarray}
The first term is the current density of the saturated matter,
and the second  current density $j^{\mu}_{val}$ shows 
the contribution from the valence nucleon as
\begin{equation}
j_{val}^{\mu} = \int \frac{d^4 p}{(2 \pi)^3} \Delta n(\vp) \delta (p_0 - \epsi_{\vp})  
\frac{ {\vPitild}_{\mu} (p)}{\Pitild_{0}(p)} .
\end{equation}
The third current density $j_{cor}^{\mu}$ is so called the core polarization 
current density, which is caused by the variation of the selfenergies of nucleons
in Fermi sea and given by
\begin{eqnarray}
j_{cor}^{\mu} &=& - 2 \sum_{\tau = \pm 1}
\int \frac{d^4 p}{(2 \pi)^3}  n_0 (\vp,\tau) 
[ \frac{\partial \Delta V(p)}{\partial p_{\mu}}  \delta [V_0 (p)]
+ \frac{\partial V_0 (p)}{\partial p_{\mu}} 
\{ \frac{\partial \delta [V (p)] }{\partial V}\}_{V=V_0} \Delta V(p) ]
\\
&=& - 2 \sum_{\tau = \pm 1}
\int \frac{d^4 p}{(2 \pi)^3}  n_0(\vp,\tau) 
\frac{\partial}{ \partial p^\mu} \{ \Delta V(p) \delta [V_0 (p)] \} .
\end{eqnarray}
Here it should be noted that the time component of the core 
polarization current density does not change the nucleon density:
\begin{equation}
j_{cor}^{0} =  - 2 \sum_{\tau = \pm 1}
\int \frac{d^4 p}{(2 \pi)^3}  n_0(\vp,\tau) 
\frac{\partial}{\partial p^0 } \{\Delta V(p) \delta [V_0 (p)] \} = 0  .
\end{equation}

Now we define the $z$-axis as the direction of the current 
at the matter.
First we calculate the isoscalar current density by taking 
the valence nucleon part of the momentum distribution to be
\begin{equation}
\Delta n(\vp,\tau) = \frac{(2 \pi)^3}{\Omega} \delta ( \vp - \va )
\end{equation}
with
\begin{equation}
\va = p_F {\hat z} ,
\end{equation}
where $\Omega$ is the volume of the system, which should be finally taken to be infinite.
The core polarization current density becomes
\begin{eqnarray}
j^3_{cor} &=& -4 \int \frac{d^4 p}{(2 \pi)^3} \theta (p_F - |\vp|)
\frac{\partial}{\partial p_z} \{ \Delta V(p) \delta [V_0 (p)] \}
\\
&=&  -4 \int \frac{d^4 p}{(2 \pi)^3} \delta (p_F - |\vp|)
\frac{p_z}{p_F} \{ \Delta V(p) \delta [V_0 (p)] \} \\
&=&  - \frac{1}{2 \pi^3}  \int d \Omega_p p_F^2 \cos{\theta_{\vp}} 
\frac{\Delta V(p)}{\Pitild_0 (p)} .
\label{jz-cor}
\end{eqnarray}

Then we separate it to several parts as
\begin{equation}
j^3_{cor} = j^3_{cor}(H) + j^3_{cor}(F)  
\end{equation}
with
\begin{eqnarray}
j^3_{cor} (H) &=& 
- \frac{1}{2 \pi^3}  \int d \Omega_p p_F^2 \cos{\theta_{\vp}} 
\frac{- \Pi^{\mu} \Delta U^{H}_{\mu} + M^{*} \Delta U^{H}_{s}
}{\Pitild_0 (p)}
 \\
j^3_{cor} (F) &=& 
- \frac{1}{2 \pi^3}  \int d \Omega_p p_F^2 \cos{\theta_{\vp}} 
\frac{- \Pi^{\mu} \Delta U^{F}_{\mu} + M^{*} \Delta U^{F}_{s} 
}{\Pitild_0 (p)} ,
\label{jzF-cor}
\end{eqnarray}
where $\Delta U^{H}_{\mu(s)}$ and $\Delta U^{F}_{\mu(s)}$ are 
shown to be contributions of $\Delta U_{\mu(s)}$.from 
Hartree and Fock parts of selfenergies, respectively.

It is not so easy to solve the above equation exactly in the RHF case
though it is possible in the RH case.
On the other hand we have known that the actual momentum dependence
is very small at least below the Fermi momentum.
Then we can suppose that a perturbative way is possible
with the respect to the momentum dependence. 

Before explaining the actual method, 
first, we would like to explain  the relativistic Hartree (RH) case. 
There the selfenergies are momentum-independent,
and the valence current density becomes
\begin{equation}
j^3_{var} = \frac{1}{\Omega} \frac{p_F}{E_F^*}.
\label{jvH}
\end{equation}
In this case  the core-polarization current density is calculated in
the following way.
\begin{eqnarray}
j^3_{cor} &=& 
- \frac{1}{2 \pi^3}  \int d \Omega_p p_F^2 \cos{\theta_{\vp}} 
\frac{ \Pi_{z} \Delta U^{H}_{z} }{E_F^*}
\nonumber \\
&=& 
- \frac{4}{3 \pi^2}\frac{p_F^3}{E_F^*} {\Delta U^{H}_{z} }
\label{jzH}
\end{eqnarray}
In the RH calculation
\begin{eqnarray}
\Delta U^{H}_z 
&=& \frac{g_v^2}{m_v^2} 
\int \frac{d^3 p}{(2 \pi)^3} n (\vp) 
\frac{p_z}{E_p^*} 
\nonumber \\
&=& \frac{g_v^2}{m_v^2} j^3 .
\label{dUH} 
\end{eqnarray}
Substituting eq.(\ref{dUH}) into eq.(\ref{jzH}), we can get 
\begin{eqnarray}
j^3 &=& \frac{1}{\Omega}\frac{p_F}{E_F^*} -
\{ \frac{g_v^2}{m_v^2} \frac{4}{3 \pi^2} \frac{p_F^3}{E_F^*} \} j^3 
\nonumber \\
&=&
\frac{1}{\Omega} \frac{p_F}{E_F^*}
\{ 1 + \frac{g_v^2}{m_v^2} \rho_B \frac{1}{E_F^*} \}^{-1}
\end{eqnarray}
In the RH case the Fermi energy is obtained as 
\begin{equation}
\epsi_F = E^*_F + \frac{g_v^2}{m_v^2} \rho_B,
\end{equation}
and then
\begin{equation}
j^3  = \frac{1}{\Omega}\frac{p_F}{\epsi_F} .
\end{equation}
In the low density region below about the saturation, $\epsi_F \approx M$,
so that we can see that the core polarization plays a role to
cancel the effect of the effective mass in the valence current density.

In the RHF case 
the contribution from the Hartree part is large, and we cannot
use the perturbative way.
Since momentum dependence of the selfenergies is not so large,
however, the difference between {$\Pitild_0$} and {$\Pi_0$} is small,
and then the Hartree part of the total current density $j^3 (H)$
can be approximately gotten with the following equation.
\begin{eqnarray}
j^3 (H) &\approx& 
\int \frac{d^3 p}{(2 \pi)^3} n(\vp) \frac{\Pi_z (p)}{\Pitild_0 (p)}
\nonumber \\
&\approx& \int \frac{d^3 p}{(2 \pi)^3} n(\vp) \frac{\Pi_z (p)}{\Pi_0 (p)}
\nonumber \\
&\approx& j^3_{var} (H) 
- \Delta U_z \int \frac{d^3 p}{(2 \pi)^3} n_0(\vp) 
\{\frac{\partial}{\partial U_z} \frac{\Pi_z(p)}{\Pi_0 (p)} \}_{\Delta U_z = 0}
\nonumber \\
&\approx& j^3_{var} (H) 
- \Delta U_z \int \frac{d^3 p}{(2 \pi)^3} n_0(\vp) 
\frac{\Pi_z(p)}{\Pi_0 (p)} \{ 1-\frac{\Pi_z^2 (p)}{\Pi_0^2 (p)} \},
\end{eqnarray}
where $j^3_{var} (H)$ is the valence part of 
the Hartree current density as
\begin{equation}
j^3_{var}(H) \approx 
\frac{1}{\Omega} \frac{\Pi_z (p_F)}{\Pitild_0 (p_F)} .
\end{equation}
 
The space component of the vector selfenergy, which is caused only by
the Fock contribution in the saturation matter, is very small, and then
$\Delta U_z$ is thought to be contributed from the Hartree parts as
\begin{equation}
\Delta U_z \approx \Delta U_z^H = \frac{g_v^2}{m_v^2} j^3 (H) .
\end{equation}
Then the Hartree contribution of the core polarization current density
 is approximately given as
\begin{equation}
j^3_{cor} (H) = j^3(H) - j^3_{var}(H) \approx 
\frac{-V_C^H (IS)}{1+V_C^H (IS)} j^3_{var}(H) 
\label{jHcor}
\end{equation} 
with
\begin{equation}
V_C^H (IS) =  \frac{g_v^2}{m_v^2} 
\int \frac{d^3 p}{(2 \pi)^3} n_0 (\vp) \frac{1}{\Pitild_0}
 \{ 1-\frac{\Pi_z^2 (p)}{\Pi_0^2 (p)} \}.
\end{equation}

As for the Fock part, the momentum dependence of selfenergies are not
so large, and its contribution is not so big in the total current density.
Instead of  getting it exactly, thus, we can use the perturbative way 
for the Fock part of the core polarization current density.
Along this line the variation of the selfenergies are taken 
to be only the contribution from the valence nucleon as
\begin{eqnarray}
\Delta U_{s}^{F} (p) & \approx &
~ \frac{3 \fpisq}{2}
\int \frac{d^3 \vk}{(2 \pi)^3} 
\Delta n({\vk}) 
\frac{M^* (k)}{ \Pitild_0 (k) } \Delta_{\pi} (p-k)   ,
\label{pUSPV}
\\
\Delta U_{\mu}^{F} (p) & \approx & 
- \frac{3 \fpisq}{2}
\int \frac{d^3 \vk}{(2 \pi)^3} 
\Delta n({\vk}) 
\frac{\Pi_{\mu} (k)}{ \Pitild_0 (k) } \Delta_{\pi} (p-k) .
\label{pUVPV}
\end{eqnarray}
Then we substitute them into the eq. (\ref{jzF-cor}), and get
\begin{equation}
j^3_{cor} (F) =
\frac{3 f_{\pi}^2 }{4 \pi^3} \tau_3 \int d \Omega_p p_F^2 \cos{\theta_{\vp}} 
\frac{\Pi_0^2 (p_F) - \Pi_v^2(p_F) + M^{*2}(p_F)}{\Pitild_0^2 (p_F)}
\Delta_\pi (0;\vp-\va) .
\label{jzFcor2}
\end{equation}

\bigskip

Next we consider the isovector current.
In the similar way we can calculate the isovector current density 
by taking the variation part of the momentum distribution as
\begin{equation}
\Delta n(\vp,\tau) = \frac{(2 \pi)^3}{\Omega} \delta ( \vp - \va ).
\end{equation}
In this work the nuclear system is taken to be the isospin symmetric 
saturated matter plus valence nucleon.
Thus, the isovector properties can be treated in the perturbative way.
Namely it can be considered that
the Dirac-fields of the valence nucleon  isoscalar one,
and that those of the nucleon in Fermi sea has a very small isovector part
coming from the valence nucleon. 

As for the Hartree part we substitute the following $V_C^H (IV)$ 
instead of $V_C^H (IS)$
into the equation (\ref{jHcor}):
\begin{equation}
V_C^H(IV) =  \frac{C_v^{IV}}{M^2} 
\int \frac{d^3 p}{(2 \pi)^3} n_0 (\vp) \frac{1}{\Pitild_0}
 \{ 1-\frac{\Pi_z^2 (p)}{\Pi_0^2 (p)} \}.
\end{equation}

As for the Fock part, furthermore, 
the variations of the selfenergies become
\begin{eqnarray}
\Delta U_{s}^{F} (p) & \approx &
 \tau_3 \frac{\fpisq}{2}
\int \frac{d^3 \vk}{(2 \pi)^3} 
\Delta n({\vk}) 
\frac{M^* (k)}{ \Pitild_0 (k) } \Delta_{\pi} (p-k)   ,
\label{pUSPVv}
\\
\Delta U_{\mu}^{F} (p) & \approx & 
- \tau_3 \frac{\fpisq}{2}
\int \frac{d^3 \vk}{(2 \pi)^3} 
\Delta n({\vk}) 
\frac{\Pi_{\mu} (k)}{ \Pitild_0 (k) } \Delta_{\pi} (p-k) .
\label{pUVPVv}
\end{eqnarray}
Then the Fock contribution of the isovector core polarization current
density is obtained as 
\begin{equation}
j^3_{cor} (F) =
\frac{ f_{\pi}^2 }{4 \pi^3}  \tau _3 \int d \Omega_p p_F^2 \cos{\theta_{\vp}} 
\frac{\Pi_0^2 (p_F) - \Pi_v^2(p_F) + M^{*2}(p_F)}{\Pitild_0^2 (p_F)}
\Delta_\pi (0;\vp-\va) .
\label{jzFcorIV}
\end{equation}

\newpage

\section{Results}

In this section we show results calculated with the above
formulation.
In this calculation we use the parameters (PF1) \cite{tomoGRF} for 
the {$\sigma$}- and {$\omega$}- exchanges to reproduce 
the saturation properties of nuclear matter: the binding energy
$BE = 16$MeV, the incompressibility $K = 200$MeV and
the effective mass $M^{*}/M = 0.7$ at the saturation
density $\rho_0 = 0.17$fm$^{-3}$.
For comparison we give results with
momentum-independent selfenergies obtained by 
the parameter-set PM1 \cite{K-con} that gives
the same saturation properties. 
As for the isovector nucleon-nucleon interaction, $C_v^{IV}$, 
we take the value of PM1.
These values are written in Table~ 1.

\begin{figure}[ht]
\vspace*{5mm}
%\begin{minipage}{15.cm}
\hspace{0.5cm}
{\includegraphics[scale=0.8]{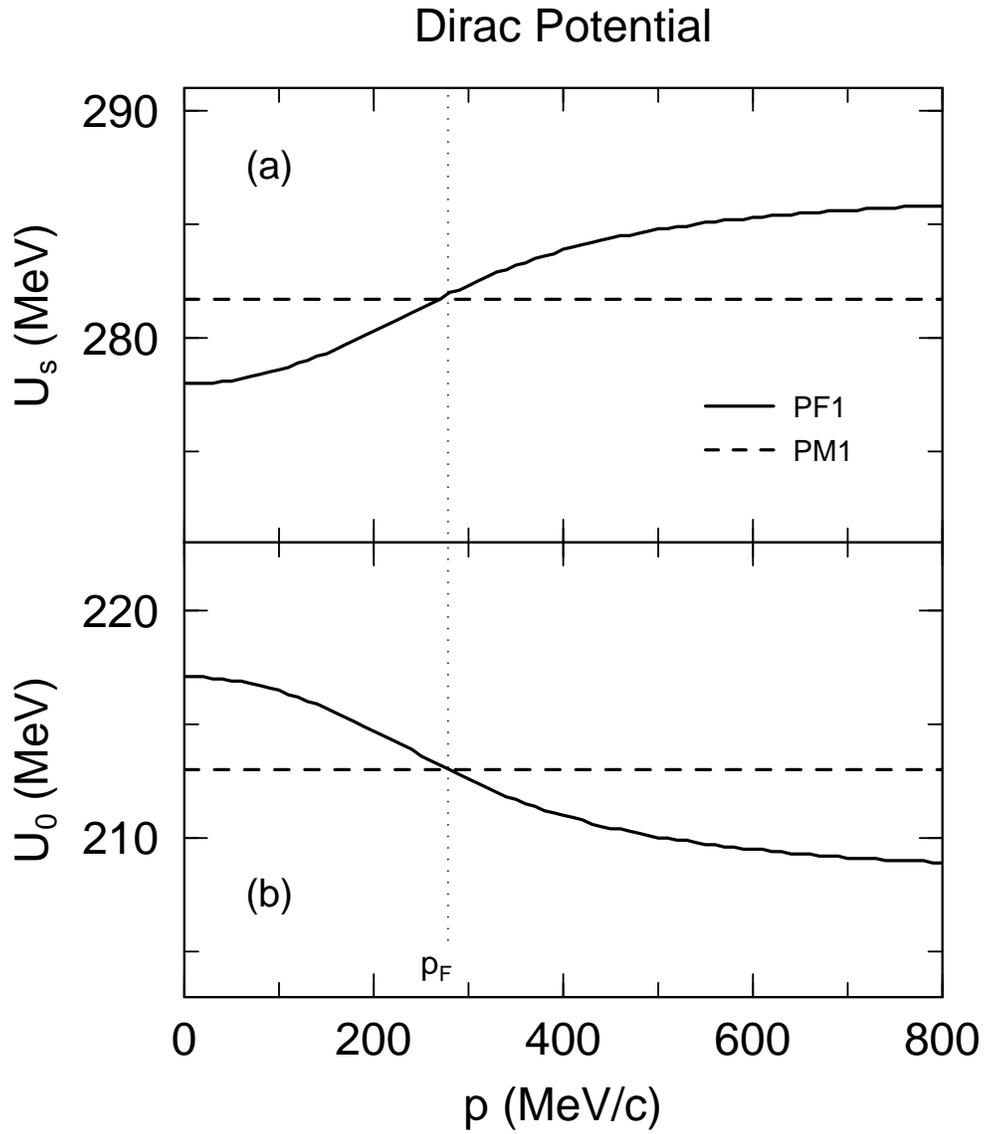}}

\caption
{\small
Momentum-dependence of the scalar (a)
and vector (b) selfenergies.
The solid  and dashed lines indicate
the results with PF1 and PM1, respectively.
The dotted line denotes the position of 
the Fermi momentum at $\rho_B = \rho_0$.}
\label{selfp}
\end{figure}

In Fig.~\ref{selfp} we draw the momentum dependence of
the scalar selfenergy $U_s (p)$ and that of  
the time component of the vector selfenergy $U_0 (p)$.
It can be seen that the variation of 
the momentum-dependent selfenergies is only 2.5 \%
at most below Fermi level, which looks very small.

\begin{figure}[ht]
\vspace*{5mm}
%\begin{minipage}{15.cm}
\hspace{0.5cm}
{\includegraphics[scale=0.8]{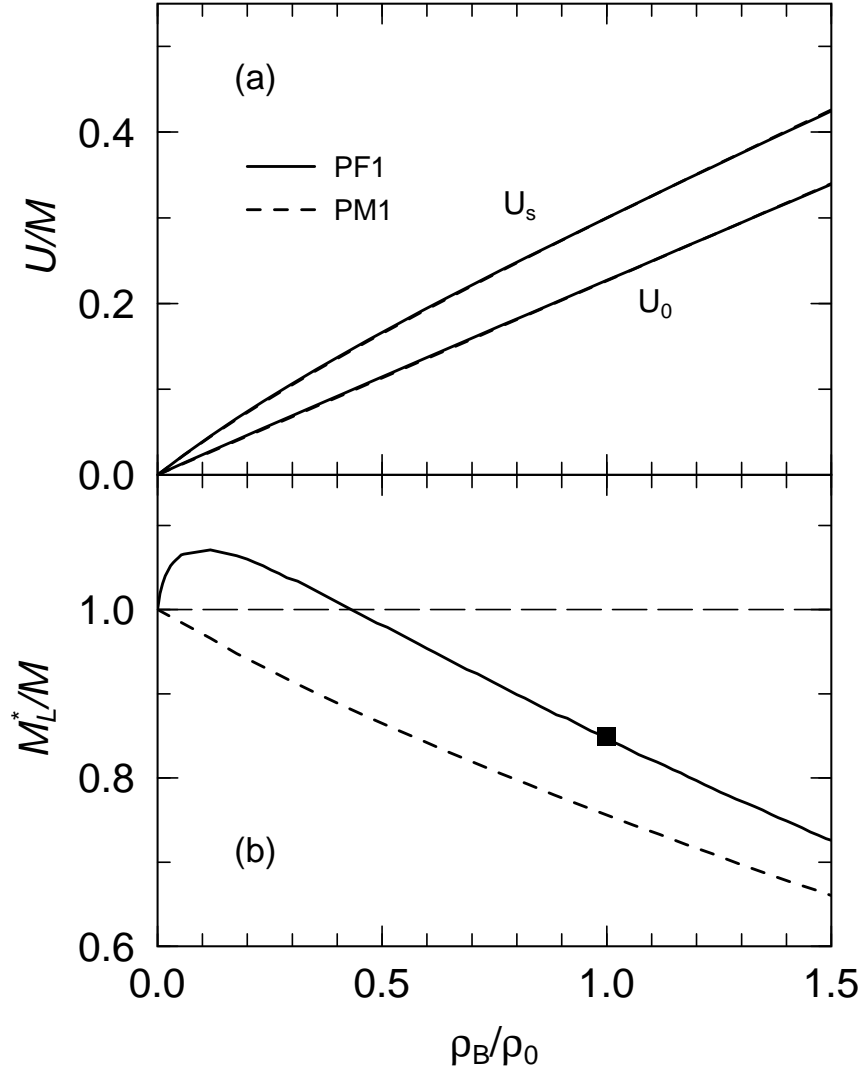}}

\caption
{\small
Density-dependence of the Dirac
selfenergies $U_s$ and $U_0$ on the Fermi-surface (a)  
and the Landau mass (b) the Landau mass (b).
The solid and dashed lines indicate the results for PF1 and PM1,
respectively, and the full square in (b) denotes the value
expected empirically from ISGQR.
}
\label{selfd}
\end{figure}

In Fig.~\ref{selfd} we show the density-dependence of the Dirac
selfenergies $U_s$ and $U_0$ on the Fermi-surface (a)  
and the Landau mass (b) with the parameter-sets, PF1 and PM1.
Though two results of $U_s$ and $U_0$ almost agree each other,
we can see rather large difference in the Landau mass: 
the value at $\rho_B = \rho_0$
is $M^{*}_L /M = 0.85$ in PF1 which is
consistent with the value expected by the analysis
of ISGQR as shown previously.  
On the contrary, the momentum-independent calculation (PM1) gives
$M^{*}_L /M = 0.74$ which overestimates the excitation energy
of ISGQR.

Hence it is shown that the very small momentum dependence 
in the nucleon selfenergies
enhances the Fermi velocity about 15 \%, and gives a 
significant difference in the Landau mass.

Furthermore we can also see an interesting behaviour of 
$M^{*}_L$ in PF1, namely, its value agrees with the bare mass at 
$\rho_B \approx 0.5 \rho_0$ and becomes larger with the decrease
of the density.
%the Landau parameter $F_1$ becomes plus in this low density region.
Effects of small Dirac effective mass
are largely cancelled at low density by the momentum dependence created 
by the one-pion exchange.

\begin{figure}[ht]
\vspace*{5mm}
%\begin{minipage}{15.cm}
\hspace{0.5cm}
{\includegraphics[scale=0.8]{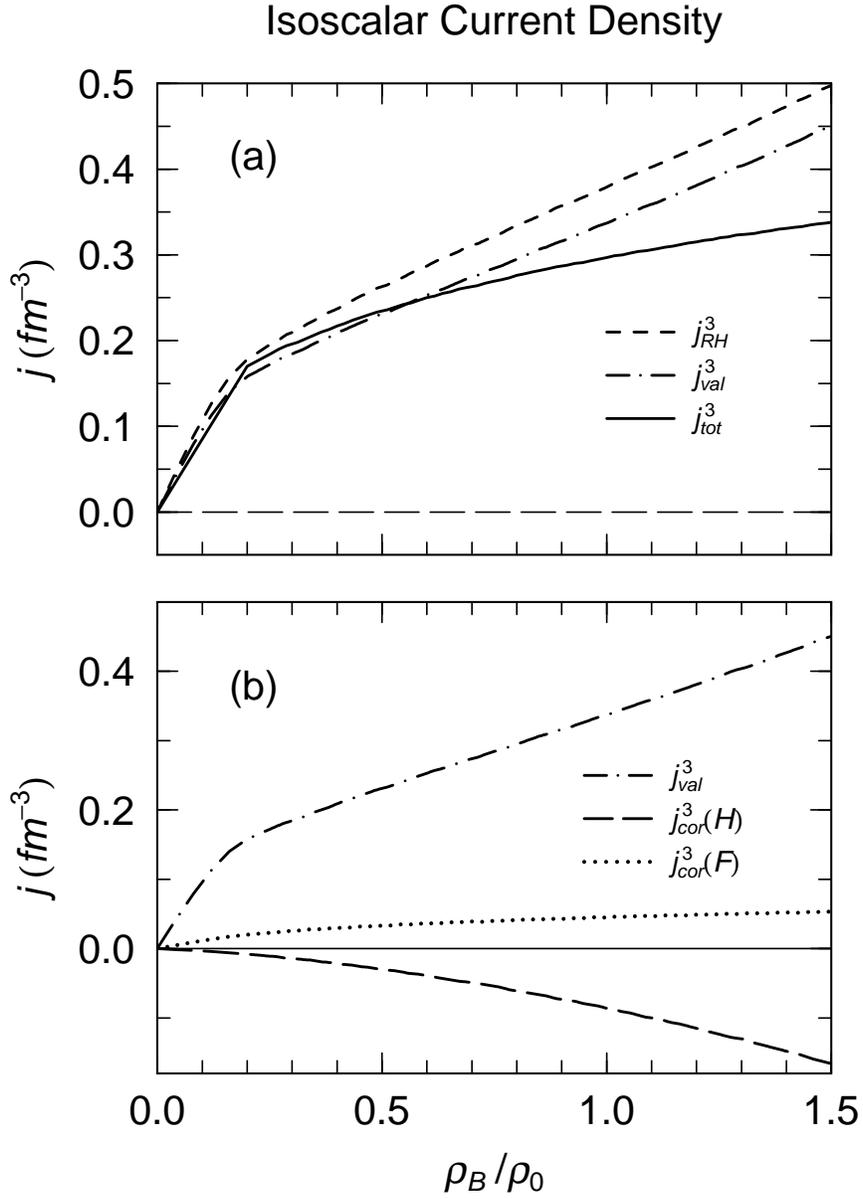}}

\caption
{\small
Density-dependence of the isoscalar nuclear current density (a)
and parts of the core polarization current density (b). 
The  chain-dotted lines indicate the valence current densities.
In the upper pannel (a) the dashed and solid lines represent
the current density for the RH approximation and 
the total current densities, respectively.
In the lower pannel (b) 
the dashed and dotted line represent
the Hartree and Fock contributions 
to the core polarization current densities, respectively.
}
\label{isosj}
\end{figure}

In Fig.~\ref{isosj} we show the density dependence of the isoscalar current density.
In the upper column (Fig.~\ref{isosj}a) the solid and chain-dotted lines
indicate the total current density
and the valence current density, respectively.
For comparison the current density for the RH approximation 
are also drawn there with the dashed line.
From that we can know that
the Fock contribution suppresses the RH current density, 
and the core polarization further suppresses it.  

In the lower column (Fig.~\ref{isosj}b) we show the contribution from the core polarization.
The long dashed and dotted lines indicate the core polarization current
density contributed
from the Hartree and Fock parts, respectively.
The Hartree contribution reduces the current density, while the Fock contribution
enhances it.

\begin{figure}[ht]
\vspace*{5mm}
%\begin{minipage}{15.cm}
\hspace{0.5cm}
{\includegraphics[scale=0.8]{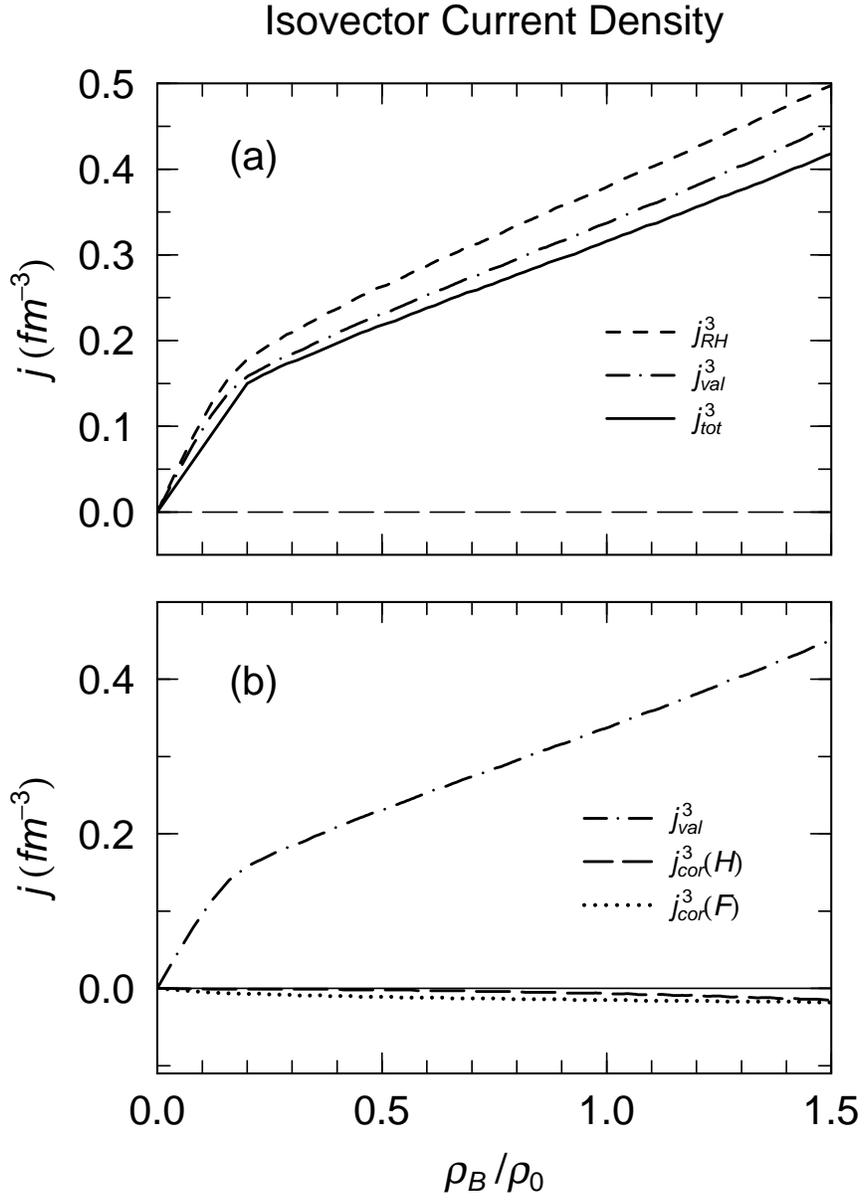}}

\caption
{\small
Density-dependence of the isovector nuclear current density (a)
and parts of the core polarization current density (b).
The meanings of lines are the same as those in Fig.~\ref{isosj}. 
}
\label{isovj}
\end{figure}

In Fig.~\ref{isovj} we show the isovector current densities; 
the meaning of each line is the same as that in Fig.~\ref{isosj}.
As for the isovector channel the core polarization does not 
affect the total current density noticeably. 

The most direct observable for the nuclear static current must be the magnetic moment;
the nuclear medium effect is examined as the discrepancy from the Schmidt value.
Thus we should compare our results with the normal current density, 
which is a current density with no medium effect and given as
\begin{equation}
j^3_o = v_F = \frac{p_F}{\epsi_F}.
\end{equation}
Here we define the following quantity as
\begin{equation}
\Delta j^3_r = \frac{j^3_{tot} - j^3_o}{j^3_o} .
\end{equation}

\begin{figure}[ht]
\vspace*{5mm}
%\begin{minipage}{15.cm}
\hspace{0.5cm}
{\includegraphics[scale=0.8]{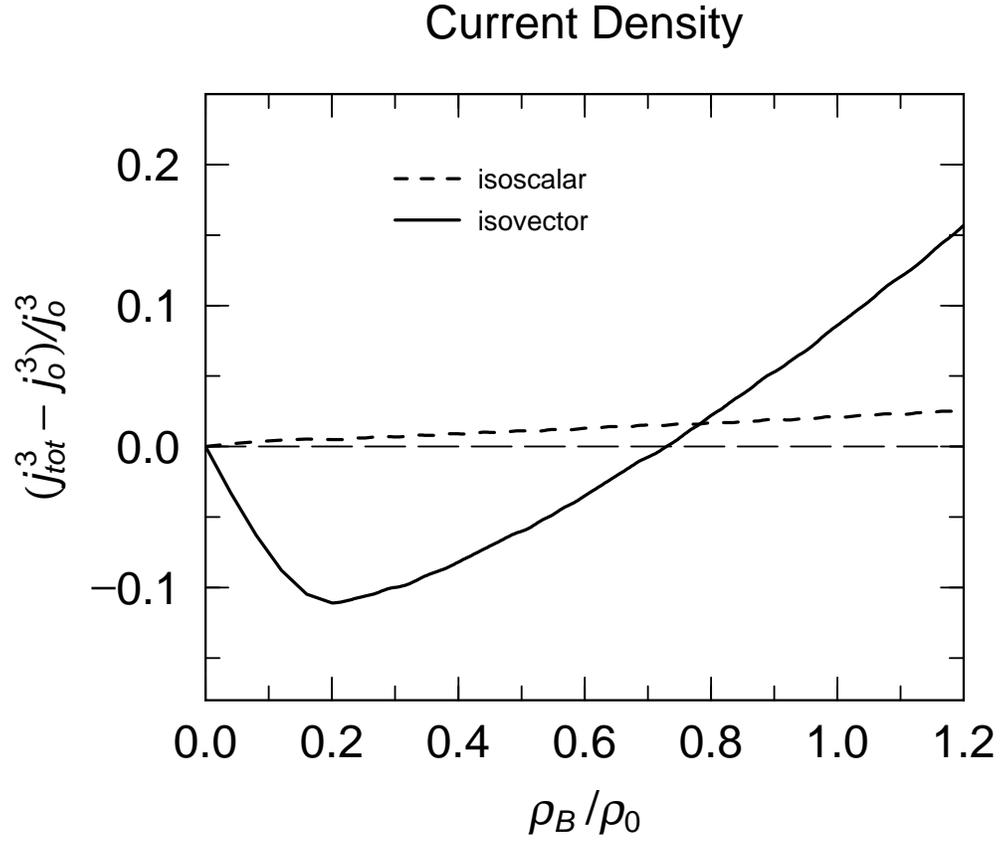}}

\caption
{\small
The density dependence of the difference between the normal current
density and
the total current density in our model, normalized by the normal current density.
The solid and dashed lines represent the isoscalar and isovector
current densities, respectively.}
\label{jcur}
\end{figure}

In Fig.~\ref{jcur} we show the density dependence of {$\Delta j^3_r$}.
As for the isoscalar current the total curren
t density, $j^3_{tot}$, 
almost agrees with the normal current density, $j^3_{o}$.
This result is consistent with the fact that the core polarization cancels
the effect caused by the effective mass.
As for the isovector, on the other hand, the core polarization does not
have significant effect.
Here we can see very interesting result that the total current density 
is 10 \% less than the normal current density 
in low density region around $\rho_B \approx \rho_0/4$.
This result is consistent with the experimental fact that the isovector magnetic
moment is 10 \% less than the Schmidt value; here we should note that the magnetic moment
indicates the medium effect in surface region.
Of course our calculation is performed for the infinite matter, and does not include
the contribution from the  anomalous part.
The result does not directly show the experimental observable, but
this results is very suggestible.

\section{Concluding Remarks}

In this paper we have defined a current operator 
which is consistent with the momentum-dependent Dirac fields.
It was shown that this current operator automatically include
the exchange current. 
This current operator describes the static spatial current 
which is determined by the Landau mass $M^{*}_L$ independently 
of the effective mass $M^{*}$.
This fact is satisfied in all cases though we here show it only 
in the one-pion exchange case which is the most effective.

Furthermore we calculate the static current in the system 
with one valence nucleon on the Fermi surface 
of the saturated nuclear matter.
In this calculation we introduce the core polarization effect.
We can confirm that the core polarization
cancels the enhancement caused by the effective mass in
the isoscalar current, but hardly affects
the isovector current.

As shown in the Ref. \cite{tomoGRF},
the very small momentum dependence in the nucleon selfenergies
enhances the Fermi velocity,
even if this momentum dependence is negligibly small for the nuclear EOS;
in the present calculation the Fermi velocity is enhanced 15 \% by 
the momentum dependence caused by the one-pion exchange.
Then we succeed to reduce the isovector current density 
in low density region;
the value of the current density is almost equivalent 
to the current density 
without effective mass at $\rho_B \approx 0.5 \rho_0$, and  
10\% suppressed around $\rho_B \approx 0.25 \rho_0$.
The latter result is consistent with that
the observed isovector magnetic moment is 10\% smaller than
the Schmidt value; of course the quantitative conclusion has not 
been so clear.   

As seen in this paper the momentum-dependent parts, which are non-local
in the finite nuclei, are very effective in observables related
with Fermi velocity 
even if these parts are small.
In future we need to discuss effects of the non-local parts of Dirac fields
to study nuclear structure and reactions. 

The typical value of effective mass is empirically known as 
$M^*_N/M_N = 0.55 - 0.7$
\cite{Hama,Tjon,DBHF,hirata,Qing}.
If we use other parameter-set which give smaller effective mass than
ours, the effects of the momentum-dependent part created by
the one-pion exchange does not have a sufficient effect
to explain the fermi velocity expected from experimental analysis.
Exchange forces of other mesons, $\sigma$, $\omega$,
$\eta$ and $\delta$, also contribute to suppressing 
the Fermi velocity if we choose the PV-coupling for $\pi$- and $\eta$-
nucleon coupling.
 
Here we should give a further comment.
Bentz et al. have shown in Ref. \cite{Bentz2} that the Landau mass 
is reduced by the one-pion exchange, which is opposite to ours.
This result is consistent with the nonrelativistic analysis
on the magnetic moment with the exchange current \cite{miyazawa}.
In this work Miyazawa showed that the exchange current enhances
the convection part, reduces the spin part, and totally reduces the
isovector magnetic moment.
In the calculation \cite{Bentz2} Bentz et al. have used 
the pseudo-scalar (PS) coupling, and the sign of 
$U^{MD}_{\mu}$ was taken to be opposite to ours.
The full HF calculation with 
the PS coupling makes too large contribution to 
the Dirac selfenergies \cite{Tjon} while
Bentz et al. calculated the Fock term with a perturbative way.
Thus a calculation with the PV coupling must be more reliable 
than that with the PS coupling.

The large discrepancy between the PS and PV coupling comes from 
relativistic effects in the one-pion exchange.
Since the pion mass is smaller than the nucleon fermi energy, 
relativistic effects must be larger in the one-pion exchange.
Miyazawa treated the one-pion exchange in the nonrelativistic way.
Thus it is not strange that our results qualitatively 
disagree with Miyazawa's one.

In this work we calculate and discuss only the convection current,
but not spin current. 
However our formalism is satisfied for the whole Dirac current, 
thus the spin-current except for the anomalous current 
must be affects by
the momentum dependence of the selfenergy in the same way. 
As for the anomalous current we have to consider the vertex correction
in another way.
Since there is no connection between the upper and lower
components
of the Dirac spinor in this current, we can suppose that 
a perturbative treatment gives sufficient results.
It is one of the future works.

Since the Fock effects are largely seen in the wide energy region 
\cite{KLW1,TOMO1}, the new current we suggested here also plays 
an important role
with the large momentum-transfer $q$. 
In future we need to discuss effects of this current in the high 
momentum transfer phenomena such as the quasielastic electron scattering 
\cite{QES}.

%References

%

\newpage

\appendix

 \section{One-Body Current Operator}

The density-dependent vertex correction $\Lambda^\mu$  is obtained as
\begin{equation}
q_\mu \Lambda^{\mu}(p+q,p) = - \Sigma(p+q) + \Sigma(p).
\label{con-c0}
\end{equation}
Within the one-boson exchange force, the Fock part of the selfenergy
is generally written in the following way.
\begin{eqnarray}
\Sigma_{F}(p) & = & i \sum_{a} C_{a}\int \frac{d^{4}k}{(2\pi)^4} 
\gamma^{a} S_N(k) \gamma_{a} {\Delta}^{(a)} (p-k) \nonumber \\
& + & i \sum_{b} {\tilde C}_{a}\int \frac{d^{4}k}{(2\pi)^4} 
[ (\psla - \ksla), \gamma^{b} ]
S_N (k) [\gamma_{b}, (\psla - \ksla)] {\Delta}^{(b)} (p-k) 
\label{Fself1}
\end{eqnarray}
where $\gamma_{a(b)}$ is the ${\gamma}$-matrix with the suffix
$a(b)$ indicating the scalar, pseudo-scalar, vector, 
axial-vector and tensor,
and $\Delta^{(a)}$ is the propagator of meson with the quantum
number indicated with the suffix $a$.

Substituting eq. (\ref{Fself1}) into eq. (\ref{con-c0}), we get
\begin{eqnarray}
q_\mu \Lambda^{\mu}(p+q,p) & = & - \Sigma(p+q) + \Sigma(p) \nonumber \\
& = & i \sum_{a} C_{a} \int \frac{d^{4}k}{(2\pi)^4} 
\gamma^{a} S_N (k) \gamma_{a}  
\{ {\Delta}^{(a)} (p-k+q) - {\Delta}^{(a)} (p-k) \} \nonumber \\
& + & 
i \sum_{b} {\tilde C}_{b} \int \frac{d^{4}k}{(2\pi)^4} 
\{ [ ({\psla}-{\ksla}+{\qsla}), \gamma^{b}] S_N (k) [ \gamma_{b}, ({\psla}-{\ksla}+{\qsla})]
 {\Delta}^{(b)} (p-k+q) \nonumber \\
& & ~~~~~~~~~~~~~~~~~ 
-  [({\psla}-{\ksla}), \gamma^{b}] S_N (k) [\gamma_{b}, (\psla-\ksla)]
 {\Delta}^{(b)} \}
\label{con-c2} 
\end{eqnarray}

When we omit the vertex form factor of the meson-nucleon coupling,
the meson propagator is given as
\begin{equation}
\Delta_{a} (k) = \frac{1}{k^2 - m_{a}^2},
\end{equation} 
and 
\begin{equation}
\Delta^{(a)} (k+q) - \Delta^{(a)}(k) = 
- \Delta^{(a)} (k+q) q(2k+q) \Delta^{(a)}(k).
\end{equation}
 
Then the above equation can be rewritten as the following expression,
\begin{eqnarray}
q^{\mu} \Lambda_{\mu} & = &  
q^{\mu} i \sum_{a} C_{a}\int \frac{d^{4}k}{(2\pi)^4} 
\gamma^{a} S_N (k) \gamma_{a} 
\{ {\Delta}^{(a)} (p-k+q) (2p - 2k + q)_{\mu}
{\Delta}^{(a)} (p-k) \} 
\nonumber \\
%%   new term
&+ & q^{\mu} i \sum_{b} {\tilde C}_{b}\int \frac{d^{4}k}{(2\pi)^4} 
\{ [\gamma^{b}, ({\psla}-{\ksla}+{\qsla})] S_N (k) [\gamma_{b}, ({\psla}-{\ksla})]
\nonumber \\
& & ~~~~~~~~~~~~~~~~~~~~~~~~~~~~
\times {\Delta}^{(b)} (p-k+q) (2p - 2k + q)_{\mu} {\Delta}^{(b)} (p-k) 
\nonumber \\ 
& & ~~~~~~~~~~~
- [ ({\psla}-{\ksla}+{\qsla}), \gamma^{b}] S_N (k)  [\gamma_{b}, \gamma_{\mu}]
{\Delta}^{(b)} (p-k+q) 
\nonumber \\
& & ~~~~~~~~~~~
- [\gamma_{\mu}, \gamma^{b}] S_N (k) [\gamma_{b}, ({\psla}-{\ksla})]
{\Delta}^{(b)} (p-k)  \}
\end{eqnarray}

From the above equation we can get the following vertex correction,
which we call $\Lambda^{(1)}$.
\begin{eqnarray}
\Lambda^{(1)}_{\mu} & = &  
i \sum_{a} C_{a}\int \frac{d^{4}k}{(2\pi)^4} 
\gamma^{a} S_N (k) \gamma_{a} 
\{ {\Delta}^{(a)} (p-k+q) (2p - 2k + q)_{\mu}
{\Delta}^{(a)} (p-k) \} 
\nonumber \\
& + & i \sum_{b} C_{b}\int \frac{d^{4}k}{(2\pi)^4} 
\{ [({\psla}-{\ksla}+{\qsla}), \gamma^{b}] S_N (k) [\gamma_{b}, ({\psla}-{\ksla})]
\nonumber \\
& & ~~~~~~~~~~~~~~~~~~~~~~~~
\times {\Delta}^{(b)} (p-k+q) (2p - 2k + q)_{\mu} {\Delta}^{(b)} (p-k) 
\nonumber \\ 
& & ~~~~~
- [({\psla}-{\ksla}+{\qsla}), \gamma^{b}] S_N (k) [\gamma_{b}, \gamma_{\mu}]
{\Delta}^{(b)} (p-k+q) 
\nonumber \\
& & ~~~~~
- [\gamma^{b},  \gamma_{\mu}] S_N (k)  [\gamma_{b}, ({\psla}-{\ksla})]
{\Delta}^{(b)} (p-k)  \}
\label{vert1} 
\end{eqnarray}

\bigskip

On the other hand the equation (\ref{Fself1}) can be rewritten as
\begin{eqnarray}
\Sigma_{F}(p) & = &i \sum_{a} C_{a}\int \frac{d^{4}k}{(2\pi)^4} 
\gamma^{a} S_N (p-k) \gamma_{a} {\Delta}^{(a)} (k) \nonumber \\
& + & i \sum_{b} {\tilde C}_{b} \int \frac{d^{4}k}{(2\pi)^4} 
[\ksla, \gamma^{b}] S_N (p-k) [\gamma_{b}, \ksla] {\Delta}^{(b)} (k) 
\label{Fself2}
\end{eqnarray}
The eq. (\ref{Fself2})  is given only by the variable transformation
($k \rightarrow p-k$) from  (\ref{Fself1}).

Substituting eq. (\ref{Fself2}) into eq. (\ref{con-c0}), then, we get
\begin{eqnarray}
q_\mu \Lambda^{\mu}(p+q,p) & = & - \Sigma(p+q) + \Sigma(p) 
\nonumber \\
& = & i \sum_{a} C_{a} \int \frac{d^{4}k}{(2\pi)^4} 
\gamma^{a} \{ S_N (p-k+q) - S_N (p-k) \} \gamma_{a}   {\Delta}^{(b)} (k) 
\nonumber \\
& + & i \sum_{b} {\tilde C}_{b} \int \frac{d^{4}k}{(2\pi)^4} 
[{\ksla}, \gamma^{b}]
\{ S_N (p-k+q) - S_N (p-k)\} [\gamma_{b}, {\ksla}] {\Delta}^{(b)} (k) 
\label{con-c3}
\end{eqnarray}

Using the WT identity (\ref{WTI}), the above equation becomes
the following expression.
\begin{eqnarray}
q^{\mu} \Lambda_{\mu} & = &  
i q^{\mu} \sum_{a} C_{a} \int \frac{d^{4}k}{(2\pi)^4} 
\gamma^{a} S_N (k+q) \Gamma_{\mu} S_N (k)  \gamma_{a}  
 {\Delta}^{(a)} (p-k) \nonumber \\
& + & i q^{\mu} \sum_{b} {\tilde C}_{b} \int \frac{d^{4}k}{(2\pi)^4} 
\{ \gamma^{b} ({\psla}-{\ksla})  S_N (k+q) \Gamma_{\mu} S(k)
 ({\psla}-{\ksla})  
\gamma_{b} {\Delta}^{(b)} (p-k)  
\label{con-c4}
\end{eqnarray}

From the above equation we can get another expression of 
the vertex correction, which we call $\Lambda^{(2)}$, as
\begin{eqnarray}
\Lambda^{(2)}_{\mu} & = &  
i \sum_{a} C_{a} \int \frac{d^{4}k}{(2\pi)^4} 
\gamma^{a} S_N (k+q) \Gamma_{\mu} S_N (k)  \gamma_{a}  
 {\Delta}^{(a)} (p-k) \nonumber \\
& + & i \sum_{b} {\tilde C}_{b} \int \frac{d^{4}k}{(2\pi)^4} 
\{ [({\psla}-{\ksla}), \gamma^{b}] S_N (k+q) \Gamma_{\mu} S(k)
[\gamma_{b}, ({\psla}-{\ksla})]  {\Delta}^{(b)} (p-k)  
\label{vert2}
\end{eqnarray}

\begin{figure}[ht]
%\vspace*{-0.5cm}
%\begin{minipage}{15.cm}
\hspace{0.2cm}
{\includegraphics[scale=0.69,angle=270]{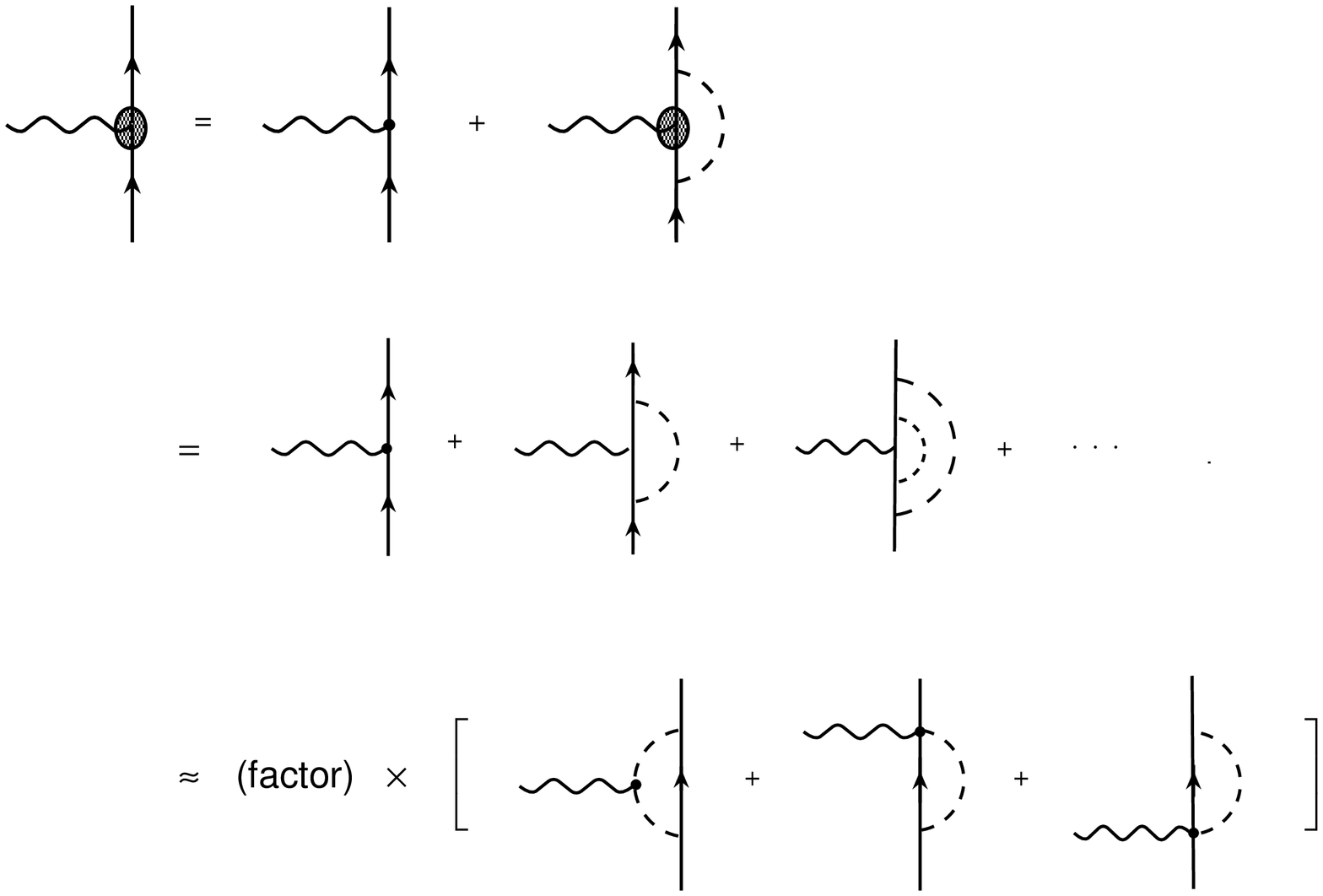}}

\caption
{\small 
Feynman diagram to show the vertex correction for the electro-magnetic
current.
The upper diagrams indicates the usual vertex correction,
and the lower ones our approximate current. 
The solid, wave and dashed lines denotes the propagators
of nucleon, meson and photon.}
\label{feyd}
\end{figure}

The Feynman diagrams corresponding to the approximate identity
$ \Lambda^{(2)}_{\mu} \approx \Lambda^{(1)}_{\mu}$
are shown in Fig.~\ref{feyd}.
The diagrams in the upper column and in the lower column indicate
the vertex corrections represented with $\Lambda^{(2)}$
and with $\Lambda^{(1)}$, respectively.
This approximation rule exhibits that 
the iteration of the vertex correction is partially equivalent
to the meson coupled with the external vector field like the meson exchange
current.

The expression of {$\Lambda^{(2)}_{\mu}$} describes
the usual vertex correction which is consistent with the Fock selfenergy.
However its calculation is not easy to be solved 
because the equation (\ref{vert2}) includes the iteration scheme. 
On the other hand we can perform the actual calculation of {$\Lambda^{(1)}_{\mu}$} 
(\ref{vert1}).
In addition we should note that the above expression is satisfied
in proton and neutron independently.
Thus we should make the following approximation rule:
\begin{eqnarray} 
&&
i \int \frac{d^{4}k}{(2\pi)^4} 
\gamma^{a} S_N (k+q) \Gamma_{\mu} S_N (k) \gamma_{a} {\Delta}^{(a)} \approx
\nonumber \\
&&
~~~~~~ i \int \frac{d^{4}k}{(2\pi)^4} 
\gamma^{a} S_N (k) \gamma_{a} \{ {\Delta}^{(a)} (p-k+q) (2p - 2k + q)_{\mu}
{\Delta}^{(a)} (p-k) \}
\label{v-rule1}
\end{eqnarray}
and
\begin{eqnarray}
&& i \int \frac{d^{4}k}{(2\pi)^4} 
\{ [({\psla}-{\ksla}), \gamma^{b}] 
S(k+q) \Gamma_{\mu} S_i (k) [\gamma^{b}, ({\psla}-{\ksla}) ] \approx
\nonumber \\
&&
i \int \frac{d^{4}k}{(2\pi)^4} 
\{[({\psla}-{\ksla}+{\qsla}), \gamma^{b}] S_N (k) [\gamma_{b}, ({\psla}-{\ksla})]
{\Delta}^{(b)} (p-k+q) (2p - 2k + q)_{\mu} {\Delta}^{(b)} (p-k) 
\nonumber \\
&&
~~~~~~~~~~~
- [({\psla}-{\ksla}+{\qsla}), \gamma^{b}] S_N(k) [\gamma_{b}, \gamma_{\mu}]
{\Delta}^{(b)} (p-k+q)
\nonumber \\
&&
~~~~~~~~~~~~~~~~~~~~~~
- [\gamma^{b},  \gamma_{\mu}] S_N(k)  [\gamma_{b}, ({\psla}-{\ksla})]
{\Delta}^{(b)} (p-k)  \} .
\label{v-rule2}
\end{eqnarray}

Here we give a comment.
If we substitute the full propagator $S_N$ into eq.(\ref{vert1}).
we have to solve the vacuum polarization, which is also very difficult.
In the usual RMF approach we usually calculate observables contributed from
the nucleon in the Fermi sea by using only the density-dependent part
$S_D$ instead of $S_F$.
In the case of the RH, where the selfenergies are momentum-independent,
the following equation is satisfied,
\begin{eqnarray}
&&i \int \frac{d^{4}k}{(2\pi)^4} 
\{ S_F(k+q) \qsla S_D(k) {\Delta}(p-k) + S_D(k) \qsla S_F (k-q) {\Delta}(p-k+q) \}
\nonumber \\
&=& i \int \frac{d^{4}k}{(2\pi)^4}  S_D(k) 
\{ {\Delta}(p-k +q) -  {\Delta}(p-k) \}.
\label{Sph}
\end{eqnarray}
This equation mentions us that we can describe particle-hole excitations
with the usual approximation that the density dependent part of the nucleon 
propagator $S_D$ (\ref{propd}) 
is used in $\Lambda^{(1)}$ instead of the full propagator. 
The actual momentum dependence is very small, and the equation (\ref{Sph})
is approximately satisfied in the RHF case, too.
From that the rules of eqs.(\ref{v-rule1}) and (\ref{v-rule2}) 
can be considered to be available under this approximation.

\newpage

\section{One-Pion Exchange Current Operator}

In this appendix we discuss details of the current operator.
As for isoscalar meson-exchanges, the correction term of
the current can be directly used in the above $\Lambda_\mu^{(2)}$
or approximately  $\Lambda_\mu^{(2)}$.
As for isovector meson exchanges, however, the current operator
includes the diagrams of the photon connect with the exchange meson 
(the mesonic current)
and that  of the photon contact with the meson-nucleon vertex 
(the contact current).

In the paper we consider only the one-pion exchange, 
which is considered to be most effective because its mass
is smallest in meson masses, and the non-locality of the Fock selfenergy
with the pion exchange is largest.

The electro-magnetic interaction Lagrangian density is describe as
\begin{equation}
{\cal L}_{em} (x)  = {\cal L}_{em}^{v} (x) +{\cal L}_{em}^{m} (x) 
                   + {\cal L}_{em}^{c} (x) 
\end{equation}
with
\begin{eqnarray}
{\cal L}_{em}^{v} & = &
 -e  {\bar \psi}(x) \gamma_{\mu} \frac{1 + \tau_3}{2} \psi ,
\label{elL1} 
\\
{\cal L}_{em}^{m} & = &
- i e \{\phi_1 (x) \partial_{\mu} \phi_2 (x) - \phi_2 (x) \partial_{\mu} \phi_1 (x) \} ,
\label{elL2} 
\\
{\cal L}_{em}^{c} & = &
- \frac{i e f_\pi}{m_\pi^2} \{ {\tilde \psi}(x) \gamma_{\mu} \gamma_{5} \tau_1 \psi (x) \phi_2 (x)
- {\tilde \psi}(x) \gamma_{\mu} \gamma_{5} \tau_2 \psi (x) \phi_1 (x) ~ \} .
\label{elL3} 
\end{eqnarray}

Now we separate the vertex corrections to three parts 
$\Lambda_\mu = \Lambda^{v}_\mu + \Lambda^{m}_\mu + \Lambda^{c}_\mu$,    
which is related with the above three parts of the electromagnetic interactions.

Here we assume that the modified Dirac current is contributed only 
from the proton as
\begin{equation}
\Gamma_\mu = \gamma_\mu \frac{ 1 + \tau_3 }{2} + \Lambda_{\mu}
= {\tilde \Gamma}_\mu \frac{ 1 + \tau_3 }{2}
\label{Jass}
\end{equation}
where $ {\tilde \Gamma}_\mu $ is an isoscalar operator.
The usual vertex correction is  given as
\begin{equation}
\Lambda_{\mu}^{v} = - \frac{i f_\pi^2}{m_{\pi}^{2}} \int \frac{d^4 k}{(2 \pi)^4}
( \psla - \ksla ) \gamma_5 \tau_a S(k+q) {\tilde \Gamma}_\mu \frac{1 + \tau_3}{2}
S(k) \tau_a \gamma_5 ( \psla - \ksla ) \Delta (p-k) 
%{\cal J}_{\mu} ( \frac{3}{2} - \frac{1}{2} \tau_3 )
\end{equation}
Then this equation can be rewritten as 
\begin{equation}
\Lambda_{\mu}^{v} = {\cal J}_{\mu} ( \frac{3}{2} - \frac{1}{2} \tau_3 )
\end{equation}
with
\begin{equation}
{\cal J}_{\mu} = \frac{i f_\pi^2}{m_{\pi}^{2}} \int \frac{d^4 k}{(2 \pi)^4}
(\psla - \ksla) \gamma_{5} S_N (p-k) {\tilde \Gamma}_{\mu} S_N (k) \gamma_5 (\psla - \ksla)
\Delta_{\pi} (p-k) .
\label{N-cur}
\end{equation}

Using the rule given by eq.(\ref{v-rule2}) the above equation is approximately
written as
\begin{eqnarray}
{\cal J}_{\mu} & \approx & {\tilde{\cal J} }_{\mu}
\nonumber \\ 
& = &  \
\frac{i f_{\pi}^2}{m_{\pi}^2} \int \frac{d^{4}k}{(2\pi)^4} 
\{ ~({\psla}-{\ksla}+{\qsla}) \gamma_{5} 
S_N (k) \gamma_{5} ({\psla}-{\ksla})
{\Delta}_{\pi} (p-k+q) (2p - 2k + q)_{\mu} {\Delta}_{\pi} (p-k) 
\nonumber \\
& & ~~~~~~~~
- ({\psla}-{\ksla}+{\qsla})\gamma_{5} S_N (k) \gamma_{5}\gamma_{\mu}
{\Delta}_{\pi} (p-k+q)
\nonumber \\
& & ~~~~~~~~
- \gamma_{5} \gamma_{\mu} S_N (k) \gamma_{5} ({\psla}-{\ksla})
{\Delta}_{\pi} (p-k)  ~ \}
\label{NA-cur}
\end{eqnarray}

Next we calculate the contribution $\Lambda^{m}$ 
from the so-called pionic current, where the photon connects with 
the pion exchange between nucleons, 
\begin{eqnarray}
\Lambda^{m}_\mu  & = &  
- \int \frac{d^4 k}{(2 \pi)^4)}
(\frac{i f_{\pi}}{m_{\pi}}) (\psla - \ksla +\qsla ) \gamma_5 \tau_i
~({\psla}-{\ksla}+{\qsla}) \gamma_{5} 
S_N (k) \gamma_{5} ({\psla}-{\ksla})
(\frac{i f_{\pi}}{m_{\pi}}) \tau_j \gamma_5 (\psla - \ksla) 
\nonumber \\
& &~~~~~~ \times
( \delta_{1 i} \delta_{2 j} - \delta_{2 i} \delta_{1 j} )
i \Delta (p-k-q) (-i) (2p -2k - q)_{\mu} i \Delta(p-k)
\nonumber \\
&=&
- \frac{ 2i f_{\pi}^2}{m_{\pi}^2} \tau_3
\int \frac{d^4 k}{(2 \pi)^4)}
\{ ~({\psla}-{\ksla}+{\qsla}) \gamma_{5} 
S_N (k) \gamma_{5} ({\psla}-{\ksla})
\nonumber \\
& &~~~~~~~~~~~~~~~~~~~~~~~~~~~~ \times
{\Delta}_{\pi} (p-k+q) (2p - 2k + q)_{\mu} {\Delta}_{\pi} (p-k) 
\label{PI-cur} 
\end{eqnarray}

\noindent
and the contribution $\Lambda^{c}$ from 
the co-called contact currents or the Siegel current 
where the photon, pion and  nucleon  connect at one vertex.
These contributions are obtained as
\begin{eqnarray}
\Lambda^{c}_\mu  & = & \
- \int \frac{d^4 k}{(2 \pi)^4)} \{~
(\frac{i f_{\pi}}{m_{\pi}}) (\psla-\ksla+\qsla) \gamma_5 \tau_i
S_N (k) (\frac{i f_{\pi}}{m_{\pi}}) \tau_j \gamma_{5} \gamma_{\mu}
{\Delta}_{\pi} (p-k+q)
\nonumber \\
&& + 
( \frac{i f_{\pi}}{m_{\pi}} )
 \gamma_\mu \gamma_5 \tau_j
S_N (p) \tau_i ( \frac{i f_{\pi}}{m_{\pi}} )
({\psla}-{\ksla}) i {\Delta}_{\pi} (p-k+q) \}
( \delta_{1 i} \delta_{2 j} - \delta_{2 i} \delta_{1 j} )
\nonumber \\
&=& \frac{- 2 i f_{\pi}^2}{m_{\pi}^2}  \tau_3
\int \frac{d^4 k}{(2 \pi)^4)} \{~
({\psla}-{\ksla}+{\qsla})\gamma_{5} S_N (k) \gamma_{5}\gamma_{\mu}
{\Delta}_{\pi} (p-k+q)
\nonumber \\
&&~~~~~~~~~~~~~~~~~~~~~
+ ~ \gamma_{5} \gamma_{\mu} S_N (k) \gamma_{5} ({\psla}-{\ksla})
{\Delta}_{\pi} (p-k)  ~ \}
\label{CN-cur}
\end{eqnarray}
From eqs.(\ref{PI-cur}) and (\ref{CN-cur}) we can obtain
\begin{equation}
\Lambda^{m}_\mu + \Lambda^{c}_\mu  =  2 \tau_3 {\tilde{\cal J} }_{\mu} .
\end{equation}

Finally the the vertex correction is given as the summation of the
above three contributions, which becomes
\begin{equation}
\Lambda_{\mu} = \Lambda_{\mu}^{v} + \Lambda_{\mu}^{m} + \Lambda_{\mu}^{c} 
= \frac{3}{2} ( 1 + \tau_3 ) \tilde{\cal J}_{\mu} .
\end{equation} 
This equation imply that the ansatz shown in eq.(\ref{Jass})
is consistent in the present calculation.

Here we would like to give some comments about the above equation.
Under the isospin symmetric matter, first, the above vertex correction satisfies 
the WT identity  as
\begin{equation}
\frac{3}{2} q^{\mu} \tilde{\cal J}_{\mu} = \Sigma_F^{\pi} (p+q) - \Sigma_F^{\pi} (p) 
\end{equation} 
with
\begin{equation}
\Sigma_F^{\pi} (p) = - U_s^F (p) + \gamma^\mu U_\mu^F (p) ,
\end{equation} 
where $U_s^F$ and $U_\mu^F$ are the Fock contributions of scalar (\ref{USPV}), 
and vector (\ref{UVPV}) selfenergies by 
the one-pion-exchange (\ref{USPV},\ref{UVPV}), respectively.

\newpage

\begin{table}[t]
\begin{center}
\begin{tabular}{|c|cccccc|}
\hline \hline 
%\stret{25pt}
 & \gs & \gv & $B_\sigma$ & $A_\sigma$ & $f_\pi$ &
 $C_v^{IV}$ \\
\hline 
%\stret{20pt}
PF1 & 9.699 & 9.880 & 27.61 & 6.134 & 1.008 & 20.32 \\ 
\hline 
%\stret{20pt}
PM1 & 9.408 & 9.993 & 23.52 & 5.651 & 0.0 & 20.32 \\
\hline
\end{tabular}

\caption
{\small Parameter sets in this paper.
In all cases have used $\mpi = 138$ MeV, \ms = 550 MeV,
$\mv = 783$ MeV and $C_\sigma$ = 0.}
\end{center}

\end{table}

\newpage

\newpage

\newpage

\newpage

\newpage

\newpage

\begin{thebibliography}{10}

\bibitem{Serot} B.D. Serot and J. D. Walecka, The relativistic Nuclear 
Many Body Problem.  In J. W. Negele and E. Vogt, editors, 
$Adv. Nucl. Phys. {\bf Vol. 16}$, page 1, Plenum Press, 1986, 
and reference therein.

\bibitem{Hama}  B. C. Clark, S. Hama, R. L. Mercer,
L. Ray and B. D. Serot, Phys. Rev. Lett. {\bf 50} (1983) 1644; \\
S. Hama, B. C. Clark, E. D. Cooper, H. S. Sherif and R. L. Mercer,
Phys. Rev. {\bf C41},   2737 (1990).

\bibitem{Tjon}
J. A. Tjon and S. J. Wallace, Phys. Rev. {\bf C36}, 1085 (1987). 

\bibitem{DBHF} W. Botermans and R. Malfliet, Phys. Rep. {\bf 198}, 115 (1990);
 \\
R. Brockmann and R. Machleidt, Phys. Rev. {\bf C42},  1965 (1990).


\bibitem{magmom} 
A. Bouyssy, S. Marcos and J.F. Mathiot, Nucl. Phys.{\bf A415}, 497 (1984).


\bibitem{Nishi} S. Nishizaki, H. Kurasawa and T. Suzuki, 
Nucl. Phys. {\bf A462}, 687 (1987).

\bibitem{Ring} H. Kurasawa and T. Suzuki, Phys. Lett. {\bf 165B}, 234 (1985).

\bibitem{Bentz}
W. Bentz, A. Arima, H. Hyuga, K. Shimizu and K. Yazaki, Nucl. Phys. {\bf A436},
 593 (1985).



\bibitem{RHF} C. J. Horowitz and B. D. Serot, Nucl. Phys. 
{\bf A399}, 529 (1983).


\bibitem{soutome} K. Soutome, T. Maruyama, K. Saito,
Nucl. Phys. {\bf 507} (1990) 731. 

\bibitem{KLW1} K. Weber, B. Bl\"attel, W. Cassing, H.-C. D\"onges,
V. Koch, A. Lang and U. Mosel, Nucl. Phys. {\bf A539},  713 (1992).

\bibitem{TOMO1} T. Maruyama, B. Bl\"attel, W. Cassing, A. Lang, U. Mosel,
K. Weber, Phys. Lett {\bf B297}, 228 (1992);\\
T. Maruyama, W. Cassing, U. Mosel, S. Teis and K. Weber, 
Nucl. Phys {\bf A552}, 571 (1994).

\bibitem{Typel}S.~Typel, nucl-th/0501056, to be published in
	Phy.~Rev.~{\bf C}.

\bibitem{tomoGRF} T. Maruyama and S. Chiba, Phys.~Rev.~{\bf C61}, 037031 (2000). 

\bibitem{K-con}
T. Maruyama, H. Fujii,  T. Muto and T. Tatsumi, 
Phys. Lett. {\bf B337}, 19 (1994); \\
H. Fujii, T. Maruyama, T. Muto and T. Tatsumi, 
Nucl. Phys. {\bf A597}, 645 (1996).

\bibitem{BM} For example, A.~Bohr and B.R.~Mottelson, ''Nuclear Structure'',
W.A.~Benjamin, Inc, and references therein.

%\bibitem{Tomo0} T. Maruyama and T. Suzuki, Phys. Lett. {\bf B219}, 43 (1989). 

\bibitem{Bentz2} W. Bentz, A. Arima and H. Baier, Nucl. Phys.
{\bf A541}, 333 (1992).

% B.Q. Chen, Z.Y. Ma, F. Gruemmer
%and S. Krewald, nucl-th 990406.

\bibitem{Vretenar}
D. Vretenar, H. Berghammer and P. Ring, Nucl. Phys. {\bf A581}
(1995) 679.

\bibitem{hirata} D. Hirata, K. Sumiyoshi, B.V. Carlson, H. Toki,
Nucl.~Phys.~{\bf A609}, 131 (1996).

\bibitem{Qing} S. Qing-biao and F. Da-chun, Phys. Rev. {\bf C43}, 2773 (1991).

\bibitem{miyazawa} H.~Miyazawa, Prog. Theor. Phys, Vol. 6, {\bf 801} (1951).

\bibitem{QES} G. Do Dang and N. Van Giai, Phys. Rev. {\bf C30}, 731 (1984),\\
S. Nishizaki, T. Maruyama, H. Kurasawa and T. Suzuki, Nucl. Phys. {\bf A485},
515 (1988).

\end{thebibliography}
\end{document}